\documentclass[a4paper]{jpconf}

\usepackage{graphicx}
\begin{document}
\title{Uniform hole doping in HgBa$_{2}$Ca$_{2}$Cu$_{3}$O$_{8+\delta}$ studied by $^{63}$Cu NMR}

\author{Yutaka Itoh$^1$, Akihiro Ogawa$^2$ and Seiji Adachi$^3$}
\address{$^1$Department of Physics, Graduate School of Science, Kyoto Sangyo University, Kamigamo-Motoyama, Kika-ku, Kyoto 603-8555, Japan}   
\address{$^2$Chugoku Electric Power Company Inc. Energia Research Institute, 3-9-1 Kagamiyama, Higashi Hiroshima, Hiroshima 739-0046, Japan} 
\address{$^3$Superconducting Sensing Technology Research Association, 2-11-19 Minowa, Kohoku, Yokohama, Kanagawa 223-0051, Japan}

\ead{yitoh@cc.kyoto-su.ac.jp}

\begin{abstract}
We studied local magnetic properties of triple-CuO$_2$-layer superconductors HgBa$_2$Ca$_2$Cu$_3$O$_{8+\delta}$ with rich $^{63}$Cu isotope (underdoped $T_\mathrm{c}$ = 124 K and optimally doped $T_\mathrm{c}$ = 134 K) by $^{63}$Cu NMR spin-echo techniques. The temperature dependences of the $^{63}$Cu nuclear spin-lattice relaxation rate $^{63}$(1/$T_1$) and the spin Knight shift $^{63}K_\mathrm{spin}$ of the inner plane Cu(1) site were nearly the same as those of the outer plane Cu(2) site in the normal states. The pseudogap temperature $T^*$ defined as the maximum temperature of $^{63}$(1/$T_1T$) was 190 K for the underdoped $T_\mathrm{c}$ = 124 sample and 143 K for the optimally doped $T_\mathrm{c}$ = 134 K sample. The NMR results indicate the uniform hole distribution in the three layers, which is consistent with a coalescence of three-derived Fermi surfaces predicted by a band theory. The room temperature in-plane $^{63}K_\mathrm{spin}^{ab}$ depends on whether the number of CuO$_2$ planes per unit cell is even or odd for the Hg-based multilayer cuprate superconductors. 
\end{abstract}

\section{Introduction}
The optimally hole doped HgBa$_2$Ca$_2$Cu$_3$O$_{8+\delta}$ (Hg1223) has the highest superconducting transition temperature $T_\mathrm{c}$ = 134 K among the ever reported cuprate superconductors~\cite{Tc1,Tc2}. 
Hg1223 has three CuO$_2$ planes per unit cell. 
Crystallographically inequivalent Cu sites are the 4-oxygen coordinated inner plane (IP) Cu(1) site and the 5-oxygen coordinated outer plane (OP) Cu(2) site. 
The left panel in Fig. 1 is an illustration of the crystal structure of Hg1223.
One of the issues on the multilayer cuprate superconductors is how the doped holes distribute in the multiple planes per unit cell. 
NMR spin-echo techniques enable us to observe selectively the inequivalent Cu sites. 
 
Transport measurements for single crystals Hg1223 with $T_\mathrm{c}$ = 134 K tell us the effects of antiferromagnetic spin fluctuations and a pseudogap~\cite{resis}. 
The pseudogap temperature of $T^{*}$ is defined as the peak temperature of the $^{63}$Cu nuclear spin-lattice relaxation rate divided by temperature 1/$T_1T$. 
The spin pseudogap opens below $T^{*}$.  
The Cu nuclear spin-lattice relaxation studies revealed the pseudogap for underdoped single crystals with $T_\mathrm{c}$ = 115 K ($T^{*}$ = 230 K)~\cite{Julien}, an underdoped Re-substituted powder sample with $T_\mathrm{c}$ = 126 K ($T^{*}$ = 172 K)~\cite{Goto}, and an optimally doped powder sample with $T_\mathrm{c}$ = 133 K ($T^{*}$ = 160 K)~\cite{Magishi,Magishi2}. 

In this paper, we report the site-selective $^{63}$Cu NMR measurements for two serial samples of the triple-CuO$_2$-layer high-$T_\mathrm{c}$ superconductors Hg1223, in which the isotope $^{63}$Cu was enriched to improve a NMR signal-to-noise ratio. 
We found uniformly hole-doped CuO$_2$ planes and an in-plane spin Knight shift $^{63}K_\mathrm{spin}^{ab}$ associated with the number parity of CuO$_2$ planes per unit cell.  

\section{Experiments} 
Polycrystalline powder samples of Hg1223 were prepared by a solid state reaction of
mixed powders of high-purity HgO, BaO, CaO, and $^{63}$Cu-rich CuO (the isotope $^{63}$Cu of concentration $>$ 90$\%$)~\cite{Fukuoka}.
The mixed powder was pelletized and encapsulated in a quartz tube under high vacuum.
The pellets sealed in the evacuated tube were fired at 775 $^{\circ}$C for 4 days.
As-synthesized samples are underdoped superconductors.
A part of the as-synthesized samples was annealed in O$_2$ gas at 370 $^{\circ}$C for 12 hours and then cooled for 10 hours.
Two serial samples with as-synthesized and O$_2$-annealed were confirmed to be in single phase by measurements of powder X-ray diffraction patterns.
The powder samples mixed in epoxy (Stycast 1266) were oriented and cured in a magnetic field of $\sim$7.5 T at room temperature.
NMR experiments were performed for the magnetically $c$-axis aligned powder samples of Hg1223.

\begin{figure}[t]
\hspace{+2pc}
\includegraphics[width=22pc]{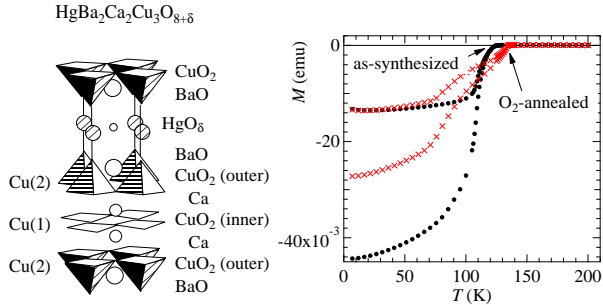}
\begin{minipage}[b]{12pc}
\caption{\label{F1}
Illustration of crystal structure of Hg1223 (left panel). 
Magnetizations of the oriented powders for a field of 100 Oe perpendicular to the aligned $c$ axis (right panel).
Closed circles and crosses are for as-synthesized and O$_2$-annealed samples, respectively. 
}
\end{minipage}
\end{figure} 
\hspace{+1pc}

The right panel in Fig.~\ref{F1} shows zero-field cooled and field cooled magnetizations of the aligned powder samples of Hg1223,
which were measured by a superconducting quantum interference device (SQUID) magnetometer (QUANTUM Design, MPMS). 
The onset temperature of superconductivity was $T_\mathrm{c}$ = 124 K for the as-synthesized underdoped sample and
$T_\mathrm{c}$ = 134 K for the O$_2$-annealed optimally doped sample~\cite{as}. 
The magnetization for the O$_2$-annealed sample shows a weak temperature dependence around 90 K.
Although the weak temperature dependence might be due to an inhomogeneous oxygen distribution,
an overdoped Hg1223 with $T_\mathrm{c}$ = 90 K is unprecedented.  

A phase-coherent-type pulsed spectrometer was utilized for the $^{63}$Cu NMR (a nuclear spin $I$ = 3/2, the nuclear gyromagnetic ratio $^{63}\gamma_n$/2$\pi$ = 11.285 MHz/T) experiments at an external magnetic field of $B_{0}$ = 7.4847 T and a reference Larmor frequency of $\nu_\mathrm{L}$ = $^{63}\gamma_nB_{0}$ = 84.465 MHz.    
The NMR frequency spectra were obtained from the integration of the $^{63}$Cu nuclear spin-echoes.   
$^{63}$Cu nuclear spin-lattice relaxation curves were obtained by an inversion recovery technique. The nuclear spin-lattice relaxation time $^{63}T_1$ was estimated by using the theoretical curve for the transition line of $I_z$ = 1/2 $\leftrightarrow$ -1/2 of a spin $I$ = 3/2. 
The subscript indices $cc$ or $ab$ of Knight shifts $^{63}K$'s and nuclear spin-lattice relaxation rates (1/$^{63}T_1$)'s indicate the direction of a static magnetic field applied along the $c$ axis or in the $ab$ plane. 

\section{Experimental results} 
\subsection{$^{63}$Cu NMR spectra}
\begin{figure}[h]
\includegraphics[width=33pc]{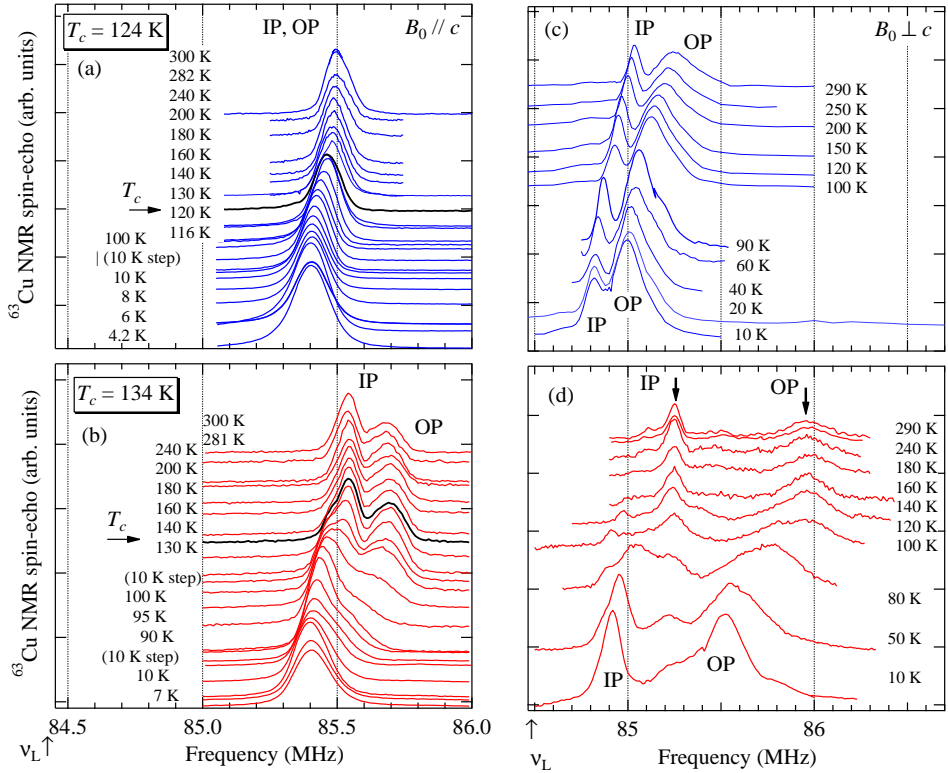}\hspace{0pc}%
\begin{minipage}[b]{38pc}\caption{\label{F2}
Central transition lines ($I_z$ = 1/2 $\leftrightarrow$ -1/2) of $^{63}$Cu NMR frequency spectra with $B_0 \parallel c$ [left panels (a) and (b)] and $B_0 \perp c$ [right panels (c) and (d)] for the as-synthesized Hg1223 [upper panels (a) and (c)] and the O$_2$-annealed Hg1223 [lower panels (b) and (d)].   
IP stands for the inner plane Cu(1) site and OP stands for the outer plane Cu(2) site. 
}
\end{minipage}
\end{figure}

Figure~\ref{F2} shows the central transition lines ($I_z$ = 1/2 $\leftrightarrow$ -1/2) of $^{63}$Cu NMR spectra with $B_0 \parallel c$ (left panels) and $B_0 \perp c$ (right panels)
for the as-synthesized Hg1223 (upper panels) and the O$_2$-annealed Hg1223 (lower panels).
Site assignment to Cu(1) and Cu(2) conforms to the previous work for a triple-layer superconductor~\cite{Bi}.
Figure~\ref{F2} shows higher orientation of the present powder samples than that in~\cite{Magishi2}. 

We focus on the central transition line ($I_z$ = 1/2 $\leftrightarrow$ -1/2) in the second order perturbation of a nuclear quadrupole interaction with uniaxial electric field gradients~\cite{Takigawa}. 
The $c$ axis is the maximum principal axis of the electric field gradient tensor. 
Then, the resonance frequency $\nu^{cc}_{\mathrm{}}$ for $B_0 \parallel c$ is expressed by the Knight shift $K^{cc}$ as
\begin{eqnarray}
\nu^{cc}_{\mathrm{}} = (1+ K^{cc})\nu_\mathrm{L}. 
\label{fcc}
\end{eqnarray} 
The resonance frequency $\nu^{ab}_{\mathrm{}}$ for $B_0 \perp c$ is expressed by the Knight shift $K^{ab}$ and the second order quadrupole shift as
\begin{eqnarray}
\nu^{ab}_{\mathrm{}} = (1+ K^{ab})\nu_\mathrm{L} + {3 \over {16}}{\nu_Q^2 \over { (1+ K^{ab})\nu_\mathrm{L}}}, 
\label{fab}
\end{eqnarray}
where $^{63}\nu_Q$ is the nuclear quadrupole resonance frequency along the maximum principal axis. 
The NQR frequencies $^{63}\nu_Q$'s were estimated to be 9.0 MHz at the inner Cu(1) and 12.0 MHz at the outer Cu(2) in the as-synthesized Hg1223, and 10.4 MHz at the inner Cu(1) and 19.5 MHz at the outer Cu(2) in the O$_2$-annealed Hg1223.

The $^{63}$Cu Knight shift $^{63}K^{\alpha\beta}$ ($\alpha\beta$ = $cc$ and $ab$) is the sum of the orbital shift $K^{\alpha\beta}_\mathrm{orb}$ and the spin shift $K^{\alpha\beta}_\mathrm{spin}$ as $K^{\alpha\beta}$ = $K^{\alpha\beta}_\mathrm{orb}$ + $K^{\alpha\beta}_\mathrm{spin}$. 
The spin shift is expressed by on-site hyperfine coupling constants $A_{\alpha\beta}$, a supertransfered coupling constant $B$, and a uniform spin susceptibility $\chi_s$ as $K^{\alpha\beta}_\mathrm{spin}$ = ($A_{\alpha\beta}$ + 4$B$)$\chi_{s}$~\cite{MR,MMP,MPS,Imai}. 

Single transition lines in Fig. 2(a) were observed both in the normal state and the superconducting state of the as-synthesized Hg1223 for $B_0 \parallel c$.
Hence, not only the spin shift $K^{cc}_\mathrm{spin}$ but also the orbital shift $K^{cc}_\mathrm{orb}$ at Cu(1) are identical to those at Cu(2). 
From the as-synthesized Hg1223 [Fig. 2(a)] to the O$_2$-annealed Hg1223 [Fig. 2(b)] for $B_0 \parallel c$,
the heat treatment with oxygen annealing leads to separate the single lines into two lines and shift them toward the higher frequency side in the normal state. 
Between 90 and 134 K in Fig. 2(b), triple Gaussian functions reproduce the NMR spectra, so that three Knight shifts for $B_0 \parallel c$ were estimated.    

Two lines in Fig. 2(c) are separated mainly due to the difference in the quadrupole shifts at Cu(1) and Cu(2) for the as-synthesized Hg1223 with $B_0 \perp c$.
Two broad lines in Fig. 2(d) are separated due to the double differences in the quadrupole shifts and the spin Knight shifts at Cu(1) and Cu(2) for the O$_2$-annealed Hg1223 with $B_0 \perp c$. 
More than two lines of Cu(2) spectra with $B_0 \perp c$ in Fig. 2(d) indicate multiple $\nu_Q$(2)'s for the O$_2$-annealed H1223,
which are consistent with multiple zero-field NQR spectra~\cite{Luders}. 
 
\subsection{$^{63}$Cu Knight shifts} 　
\begin{figure}[h]
\begin{minipage}{18pc}
\includegraphics[width=18pc]{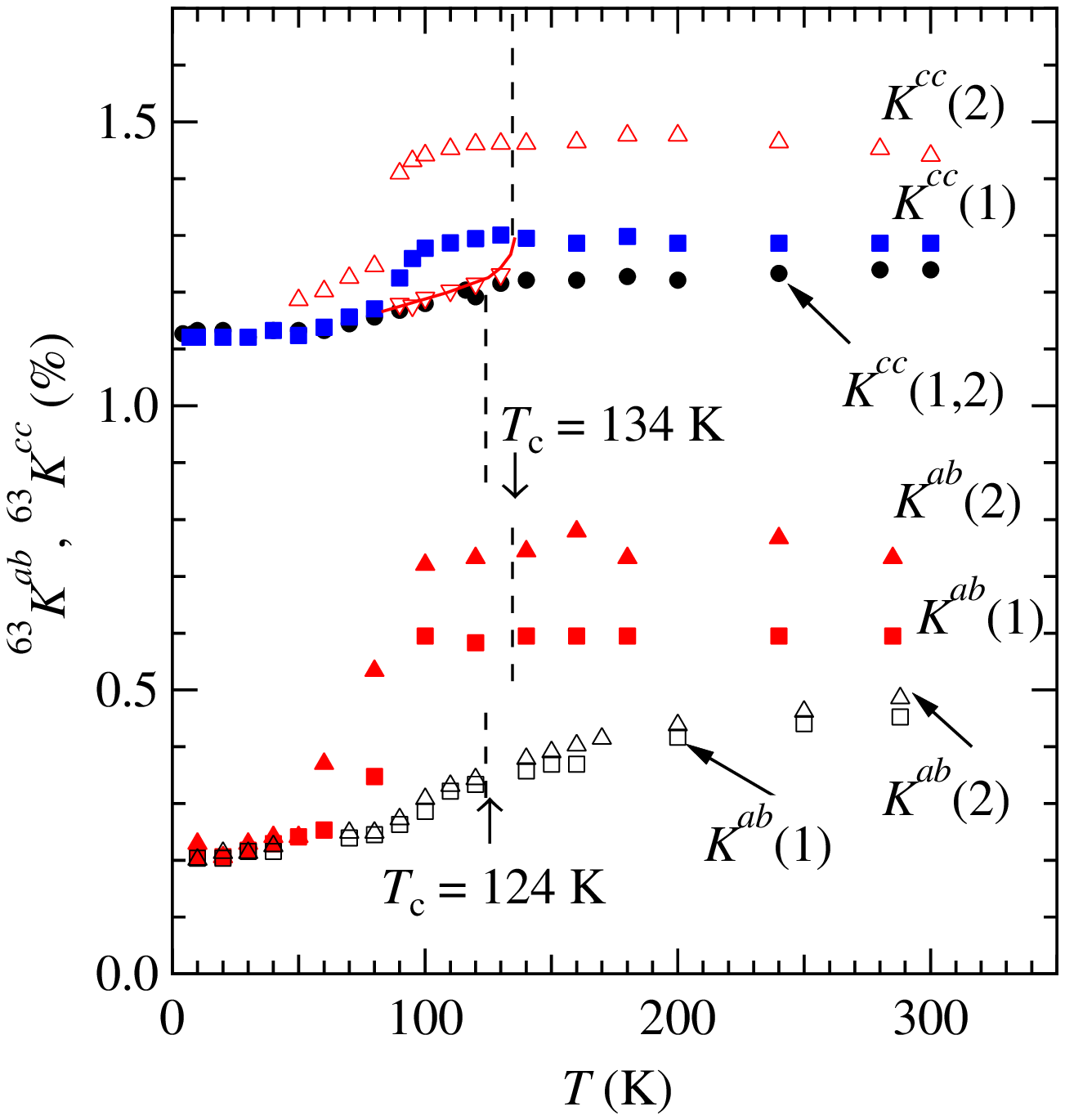}
\caption{\label{F3}
$^{63}$Cu Knight shifts $^{63}K^{cc, ab}$(1, 2) for Hg1223 from underdoped (black symbols) to optimally doped (red and blue symbols).  
Squares and upward triangles indicate the inner plane Cu(1) site and the outer plane Cu(2) site, respectively. 
Downward open triangles at 90$-$134 K are the lowest frequency component in triple Gaussian fitting functions for the O$_2$-annealed Hg1223 with $B_0 \parallel c$. 
}
\end{minipage}\hspace{2pc}%
\begin{minipage}{18pc}
\includegraphics[width=18pc]{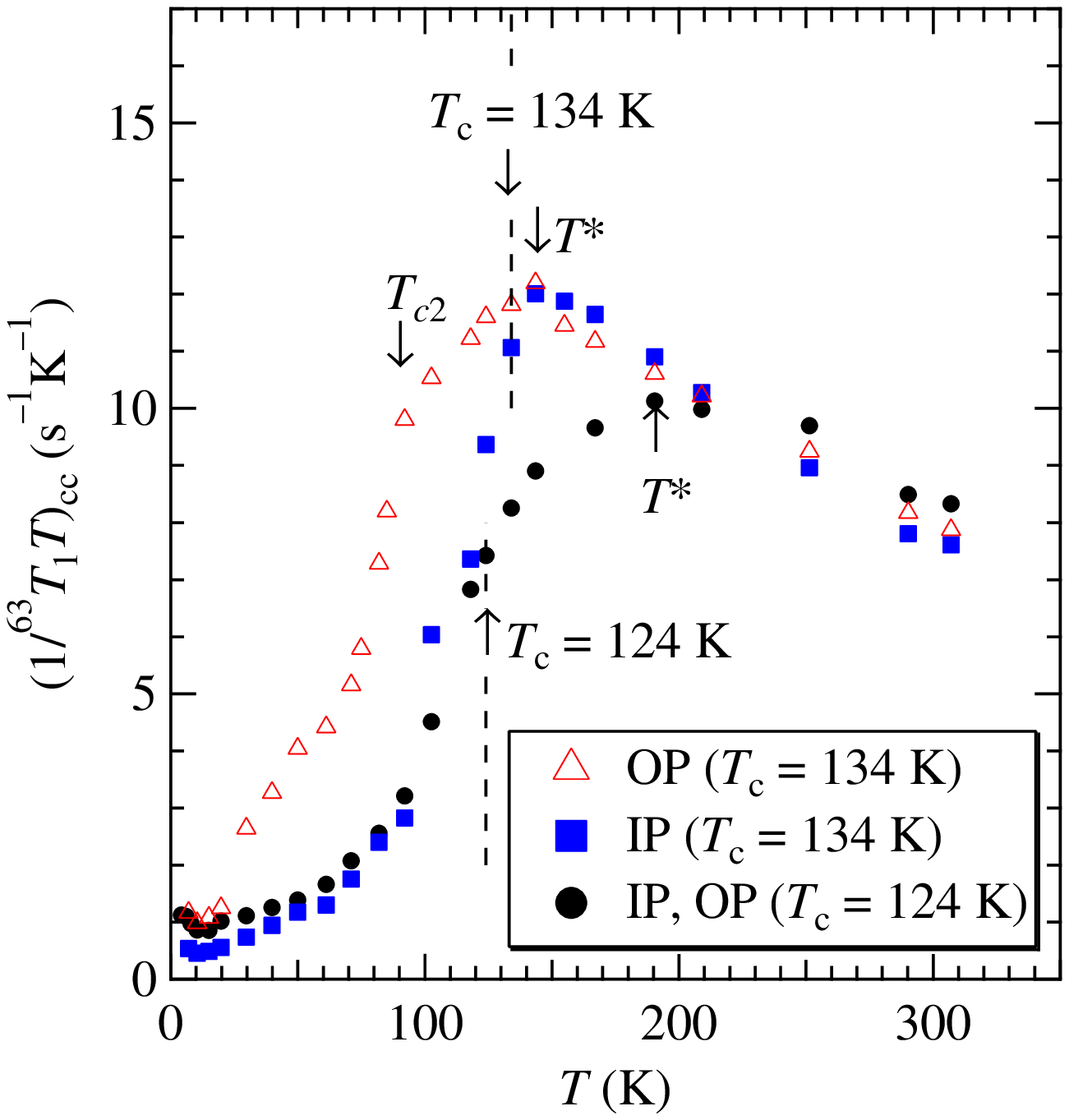}
\caption{\label{F4}
$^{63}$Cu nuclear spin-lattice relaxation rates divided by temperature (1/$^{63}T_1T$)$_{cc}$ with $B_0 \parallel c$ for Hg1223 from underdoped to optimally doped.  
Black symbols are for the as-synthesized Hg1223. 
Red and blue symbols are for the O$_2$-annealed Hg1223.  
Squares and upward triangles indicate the inner plane Cu(1) site and the outer plane Cu(2) site, respectively. 
}
\end{minipage} 
\end{figure}

Figure~\ref{F3} shows $^{63}$Cu Knight shifts $^{63}K^{\alpha\beta}$(1, 2) ($\alpha\beta$ = $cc$ and $ab$) in Hg1223 from underdoped to optimally doped, which were estimated from Fig. 2 by using~(\ref{fcc}) and~(\ref{fab}).   
For the as-synthesized Hg1223, $K^{\alpha\beta}$(1) is nearly the same as $K^{\alpha\beta}$(2). 
All the Knight shifts $K^{\alpha\beta}$(1, 2) decrease with cooling in the normal state, which indicates the opening of a large pseudogap.  
We estimated $\Delta K_\mathrm{spin}^{cc}/\Delta K_\mathrm{spin}^{ab}$ = 0.42. 
The inner plane spin susceptibility $\chi_s$(1) is nearly the same as the outer plane $\chi_s$(2).  
The distribution of the doped holes is uniform in the three layers. 
The uniform hole distribution in the three layers is consistent with a coalescence of the three-derived Fermi surfaces predicted by a band theory~\cite{band}.
The uniform hole distribution in the inequivalent CuO$_2$ planes has also been found in the previously reported Hg1223~\cite{Julien,Magishi} and the underdoped triple-layer Tl$_2$Ba$_2$Ca$_2$Cu$_3$O$_{10-\delta}$~\cite{TL2223}. 

For the O$_2$-annealed Hg1223, all the Knight shifts $K^{\alpha\beta}$(1, 2) are nearly independent of temperature in the normal state and decrease below 90 K but not 134 K. 
The differences between $K^{\alpha\beta}$(1) and $K^{\alpha\beta}$(2) in the normal state are nearly independent of temperature above 134 K.  
The temperature independent spin susceptibility is typical for the optimally and overdoped superconductors.  
The lowest frequency component in the triple Gaussian functions for $B_0 \parallel c$ starts to decrease below 134 K. 

\begin{figure}[t]
\includegraphics[width=17pc]{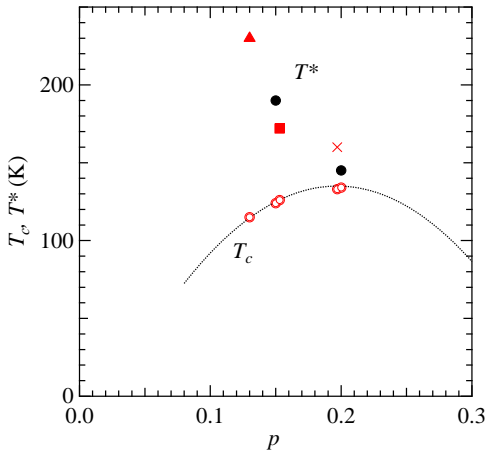}\hspace{0pc}%
\begin{minipage}[b]{20pc}
\caption{\label{F5}
Phase diagram as a function of the hole concentration $p_{}$ for Hg1223.  
The hole concentration $p$ is estimated from~\cite{Fukuoka}. 
A triangle~\cite{Julien}, a cross~\cite{Magishi}, a square~\cite{Goto}, and closed circles (the present works) indicate pseudogap temperatures $T^*$'s. 
Open circles are $T_\mathrm{c}$'s. A dotted curve is guide for the eye. 
}
\end{minipage}
\end{figure} 
We consider three possible reasons why no drop at 134 K in $K^{cc, ab}$(1,2) but the drop at 90 K is found for the O$_2$-annealed Hg1223, (i) the inhomogeneous oxygen distribution with overdoped $T_{c2}$ = 90 K, (ii) coexistence of vortex solid and liquid states~\cite{Reyes,ItohTL2201}, and (iii) bifurcation of superconducting order parameters in the inner and outer CuO$_2$ planes~\cite{2band}.
A slightly overdoped Hg1223 is known to be synthesized~\cite{Fukuoka}. 
An overdoped $T_\mathrm{c2}$ = 90 K could be the first in Hg1223. 
Irreversibility lines of Hg1223 are lower than those of Tl-layer cuprate superconductors~\cite{Rav}.
Then, a vortex liquid state may be broad in a magnetic filed vs temperature diagram of Hg1223.
Different superconducting order parameters at Cu(1) and Cu(2) could result from a magnetic bilayer coupling.  

\subsection{$^{63}$Cu nuclear spin-lattice relaxation rates divided by temperature} 　
Figure~\ref{F4} shows $^{63}$Cu nuclear spin-lattice relaxation rates divided by temperature (1/$^{63}T_1T$)$_{cc}$ with $B_0 \parallel c$ for Hg1223 from underdoped to optimally doped. 
For the as-synthesized Hg1223, (1/$^{63}T_1T$)$_{cc}$ increases with decreasing temperature and starts to decreases at the pseudogap temperature $T^*$ = 190 K. 
For the O$_2$-annealed Hg1223, both (1/$^{63}T_1T$)$_{cc}$'s at Cu(1) and Cu(2) increase with decreasing temperature in nearly the same Curie-Weiss temperature dependence and start to decrease separately at $T^*$ = 143 K.  
Below $T^*$ = 143 K and in the superconducting state below 134 K, the temperature dependence in (1/$^{63}T_1T$)$_{cc}$ of Cu(2) is different from that of Cu(1). 
(1/$^{63}T_1T$)$_{cc}$ of Cu(2) shows a secondary rapid decrease below $T_{c2}$ = 90 K, which may be associated with the drop of the Knight shift at 90 K. 
Nearly the same $T$ dependences above 143 K and the large difference below 134 K at Cu(1) and Cu(2) have been found in (1/$^{63}T_1T$)$_{ab}$ for $B_0 \perp c$. 
(1/$T_1)_{ab}$/(1/$T_1)_{cc}\sim$ 1.7 around 200 K was estimated for both samples.   

Figure~\ref{F5} shows a phase diagram of Hg1223 with respect to the hole concentration $p$~\cite{Fukuoka}. 
The pseudogap temperatures $T^*$'s and $T_\mathrm{c}$'s are plotted against the hole concentration $p_{}$.  
$T^*$ monotonically decreases as $p$ increases both for Cu(1) and Cu(2) in Hg1223.
The uniform pseudogap temperatures $T^*$'s characterize Hg1223. 

\section{Discussions} 
\subsection{Even and odd number of CuO$_2$ planes per unit cell}
\begin{figure}[t]
\begin{minipage}{17pc}
\includegraphics[width=19pc]{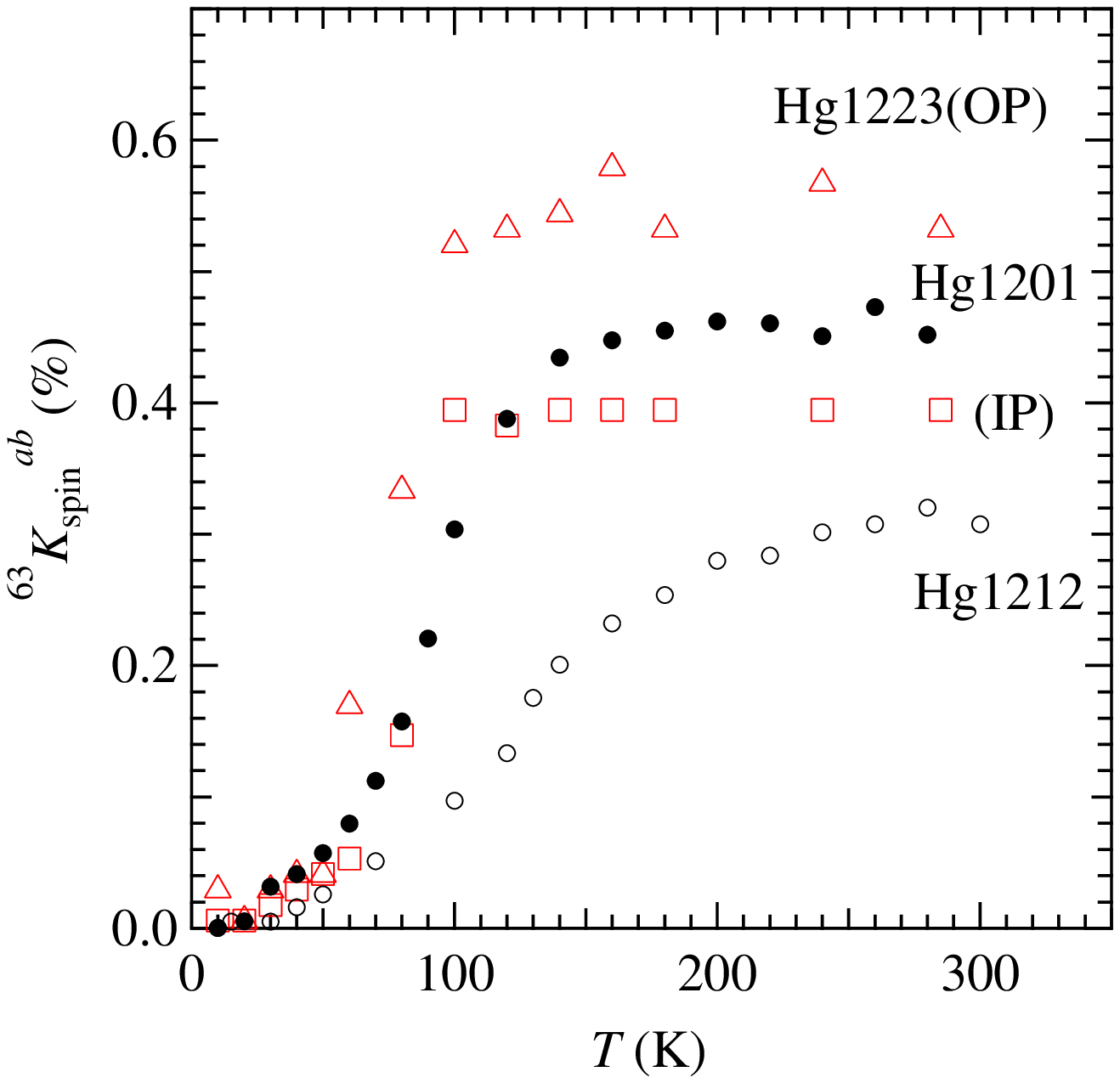}
\caption{\label{F6}
$T$ dependences of $^{63}$Cu spin Knight shifts $^{63}K_\mathrm{spin}^{ab}$'s for a single-layer HaBa$_2$CuO$_{4+\delta}$ (Hg1201) with $T_\mathrm{c}$ = 98 K~\cite{ItohHg1201}, a double-layer HaBa$_2$Ca$_{}$Cu$_2$O$_{6+\delta}$ (Hg1212) with $T_\mathrm{c}$ = 127 K~\cite{ItohHg1212}, and the present triple-layer Hg1223 with $T_\mathrm{c}$ = 134 K at the optimally doping level.  
}
\end{minipage}\hspace{2pc}%
\begin{minipage}{19pc}
\includegraphics[width=20pc]{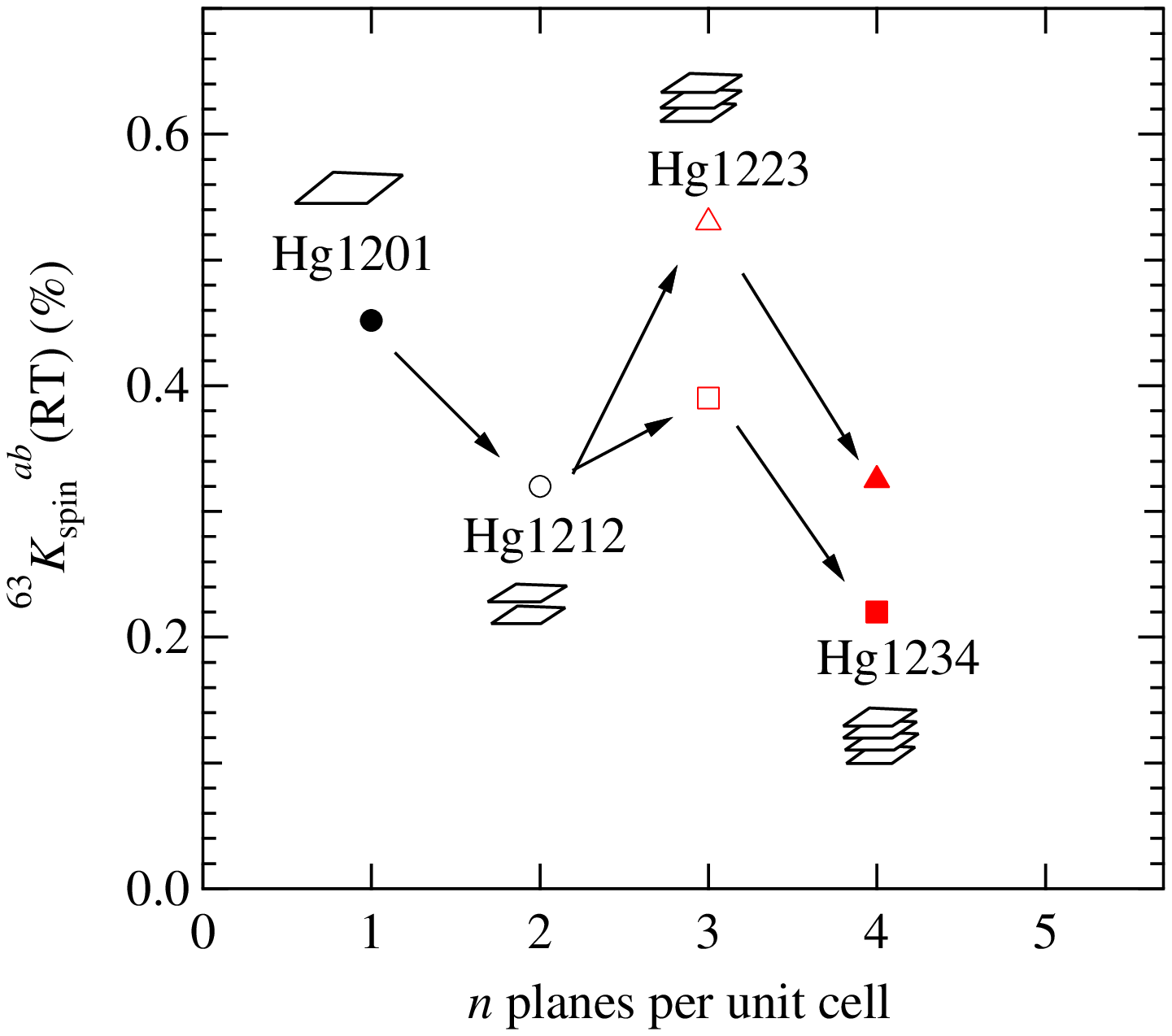}
\caption{\label{F7}
$^{63}$Cu spin Knight shifts $^{63}K_\mathrm{spin}^{ab}$'s (280$-$300 K) at the optimally doping level plotted against the number $n$ of CuO$_2$ planes per unit cell. 
Data for Hg1234~\cite{Tokunaga} are included. 
Parallelograms illustrate the number parity of CuO$_2$ planes per unit cell.
Arrows are guide for the eye. 
}
\end{minipage} 
\end{figure}

Figure~\ref{F6} shows temperature evolution of $^{63}$Cu spin Knight shifts $^{63}K_\mathrm{spin}^{ab}$'s for a single-layer HaBa$_2$CuO$_{4+\delta}$ (Hg1201) with $T_\mathrm{c}$ = 98 K~\cite{ItohHg1201}, a double-layer HaBa$_2$Ca$_{}$Cu$_2$O$_{6+\delta}$ (Hg1212) with $T_\mathrm{c}$ = 127 K~\cite{ItohHg1212}, and the present triple-layer Hg1223 with $T_\mathrm{c}$ = 134 K at the optimally doping level.  
$^{63}K_\mathrm{spin}^{ab}$ of Hg1212 is smaller than that of Hg1201. 
$^{63}K_\mathrm{spin}^{ab}$ of Hg1223 is larger than that of Hg1212. 
Figure~\ref{F6} indicates that the value of the $^{63}$Cu spin Knight shift $^{63}K_\mathrm{spin}^{ab}$ at 280$-$300 K depends on whether the number of CuO$_2$ planes per unit cell is even or odd.  

Figure~\ref{F7} shows the room temperature $^{63}$Cu spin Knight shifts $^{63}K_\mathrm{spin}^{ab}$(RT) at the optimally doping level against the number $n$ of CuO$_2$ planes per unit cell. 
The values $^{63}K_\mathrm{spin}^{ab}$'s at 280$-$300 K are adopted as the room temperature shifts. 
Data for optimally doped Hg1234 (HgBa$_2$Ca$_3$Cu$_4$O$_{10+\delta}$) are from~\cite{Tokunaga}. 
The relation of $^{63}K_\mathrm{spin}^{ab}$(RT) vs $p$ for the single layers is different from that for the double layers~\cite{ItohNum},
which breaks a single universal relation of $^{63}K_\mathrm{spin}^{ab}$(RT) vs $p$ in~\cite{Mukuda}.
The value of $^{63}K_\mathrm{spin}^{ab}$(RT) for the even number of CuO$_2$ planes per unit cell is smaller than that for the odd number of CuO$_2$ planes per unit cell.  
The number parity of the CuO$_2$ planes plays a significant role in $^{63}K_\mathrm{spin}^{ab}$(RT).

The value of $^{63}K_\mathrm{spin}^{ab}$(RT) at the optimally doped level tends to decrease as the number $n$ of the CuO$_2$ planes increases.
All the inner and outer $^{63}K_\mathrm{spin}^{ab}$(RT)'s in an optimally doped five-layer superconductor Hg1245~\cite{Mukuda5} are larger than those in the four-layer Hg1234~\cite{Tokunaga}, but the single $^{63}K_\mathrm{spin}^{ab}$(RT) for the two inner Cu sites in the five-layer Hg1245 is smaller than that in the double-layer Hg1212~\cite{ItohHg1212}.  

An intrabilayer coupling is no chief ingredient for the pseudogap, because the pseduogap is in the single layer Hg1201~\cite{ItohHg1201,ItohHg12010}. 
However, the intrabilayer superexchange coupling constant $J_b$ between the double CuO$_2$ planes in antiferromagnetic insulators YBa$_2$Cu$_3$O$_{6+\delta}$ is estimated to be $J_b$ $\sim$ 0.1$J_{\parallel}$ with an in-plane superexchange coupling constant $J_{\parallel}$ from observation of an optical magnon~\cite{Jperp,Jperp1,Jperp2}. 
The intrabilayer coupling may induce spin pairing between adjacent CuO$_2$ planes and then may enhance the spin pseudogap,
so that the uniform spin susceptibility of the even number of CuO$_2$ layers may be smaller than that of the odd number of CuO$_2$ layers~\cite{MM,Norman,JK,Kishine}. 

The O$_2$-annealed Hg1223 shows $^{63}K_\mathrm{spin}^{\alpha\beta}$(2) $>$ $^{63}K_\mathrm{spin}^{\alpha\beta}$(1) in the normal state. 
We consider two possible reasons for their difference with the uniform hole distribution. 
One is a constant second spin susceptibility, which was inferred from anisotropic Knight shifts of single crystals Hg1201~\cite{Jurugen1,Jurugen2}.
Such a second spin susceptibility of Cu(2) may be different from that of Cu(1). 
The other is the intrabilayer coupling effect with a doping dependent $J_b$.
The number of the nearest neighbor CuO$_2$ plane is 2 at Cu(1) and 1 at Cu(2). 
Then, we speculate the mean field expressions of
$\chi_{s}(1) = \chi_{s}/(1+2J_{b}\chi_s)$ and $\chi_{s}(2) = \chi_{s}/(1+J_{b}\chi_s)$ 
which lead to $\chi_s$(2) $> \chi_s$(1). 
 
\subsection{$^{63}$Cu hyperfine coupling constant ratios}
The ratios of the $^{63}$Cu hyperfine coupling constants $A_{cc}$, $A_{ab}$, and $B$ are expressed by the experimental anisotropies $r_u$ = $\Delta K_\mathrm{spin}^{cc}/\Delta K_\mathrm{spin}^{ab}$ and $r_A$ = $\sqrt{2r_{AF} - 1}$ [$r_{AF}$ = (1/$T_1)_{ab}$/(1/$T_1)_{cc}$]~\cite{ItohHF} as
\begin{eqnarray} 
{A_{cc} \over {4B}} \approx  - {{r_{A}+r_{u} - 2r_{A}r_{u}} \over {r_{A}-r_{u}}}\mathrm{,\quad}
{A_{ab} \over {4B}} \approx {{r_{A}+r_{u} - 2} \over {r_{A}-r_{u}}}, 
\label{AccAab}
\end{eqnarray}
and
\begin{eqnarray} 
{A_{ab} \over {A_{cc}}} \approx  - {{r_{A}+r_{u} - 2} \over {r_{A}+r_{u} - 2r_{A}r_{u}}}. 
\label{Accab}
\end{eqnarray}

Substituting $r_u$ = 0.42 and $r_{AF}$ = 1.7 into~(\ref{AccAab}) and~(\ref{Accab}), we estimated the ratios of the $^{63}$Cu hyperfine coupling constants for the as-synthesized Hg1223 as $A_{cc}$/4$B$ = $-$0.59, $A_{ab}$/4$B$ = $-$0.03 and $A_{ab}$/$A_{cc}$ = +0.05.
Assuming $A_{ab}$ + 4$B$ = 182 kOe/$\mu_\mathrm{B}$~\cite{ItohHF}, we obtained the values of $A_{cc}$ = $-$110, $A_{ab}$ = $-$5.6, and $B$ = 47 kOe/$\mu_B$ for the underdoped Hg1223.  
Cancellation of a relatively large core polarization and a spin dipole field from the 3$d_{x^2-y^2}$ orbital can result in the small negative $A_{ab}$~\cite{Bleaney}. 

\section{Conclusions}
The site-selective $^{63}$Cu NMR results indicate uniform hole distribution in the three CuO$_2$ planes of Hg1223 near the optimally and less doping levels.  
For the Hg-based multilayer cuprate superconductors at the optimally doping level, the room temperature spin shift $^{63}K_\mathrm{spin}^{ab}$ in the even number of CuO$_2$ planes per unit cell is smaller than that in the odd number of CuO$_2$ planes per unit cell. 


\section*{References}

\smallskip
 


\medskip
\begin{thebibliography}{9}

\bibitem{Tc1}Schilling A, Cautoni M, Guo J D and H.R. Ott H R 1993 {\it Nature} {\bf 363} 56
\bibitem{Tc2}Antipov E V, Loureiro S M, Chaillout C, Capponi J J, Bordet P, Tholence J L, Putilin S N and Marezio M 1993 {\it Physica} C {\bf 215} 1
\bibitem{resis}Carrington A, Colson D, Dumont Y, Ayache C, Bertinotti A, Marucco J F 1994 {\it Physica} C {\bf 234} 1
\bibitem{Julien}Julien M H, Carretta P, Horvati\'{c} M, Berthier C, Berthier Y, S\'{e}gransan P, Carrington A and Colson D 1996 {\it Phys. Rev. Lett.} {\bf 76} 4238 
\bibitem{Goto}Goto A, Shimizu T, Sastry P V P S S and Schwartz J 1999 {\it Phys. Rev.} B {\bf 59} 14169 
\bibitem{Magishi}Magishi K, Kitaoka Y, Zheng G q, Asayama K, Tokiwa K, Iyo A and Ihara H 1995 {\it J. Phys. Soc. Jpn.} {\bf 64} 4561
\bibitem{Magishi2}Magishi K, Kitaoka Y, Zheng G q, Asayama K, Tokiwa K, Iyo A and Ihara H 1996 {\it Phys. Rev.} {\bf 53} R8906
\bibitem{Fukuoka}Fukuoka A, Tokiwa-Yamamoto A, Itoh M, Usami R, Adachi S and Tanabe K 1997 {\it Phys. Rev.} {\bf 55} 6612
\bibitem{as}Colson D, Bertinotti A, Hammann J, Marucco J F and Pinatel A 1994 {\it Physica} C {\bf 233} 231
\bibitem{Bi}Statt B W and Song L M 1993 {\it Phys. Rev.} {\bf 48} 3536
\bibitem{Takigawa}Takigawa M, Hammel P C, Heffner R H, Fisk Z, Smith J L and Schwarz R B 1989 {\it Phys. Rev.} B {\bf 39}  300
\bibitem{MR}Mila F and Rice T M 1989 {\it Physica} C {\bf 157} 561 
\bibitem{MMP}Millis A J, Monien H and Pines D 1990 {\it Phys. Rev.} B {\bf42} 167
\bibitem{MPS}Monien H, Pines D and Slichter C P 1990 {\it Phys. Rev.} B {\bf 41} 11120  
\bibitem{Imai}Imai T 1990 {\it J. Phys. Soc. Jpn.} {\bf 59} 2508 
\bibitem{Luders}Breitzke H, Eremin I, Manske D, Antipov E V, L\"{u}ders K 2004 {\it Physica} C {\bf 406} 27
\bibitem{band}Singh D J and Pickett W E 1994 {\it Phys. Rev. Lett.} {\bf 73} 476
\bibitem{TL2223}Piskunov Y V, Mikhalev K N, Zhdanov Yu I, Gerashenko A P, Verkhovskii S V, Okulova K A, Medvedev E Yu, Yakubovskii A Yu, Shustov L D, Bellot P V and Trokiner A 1998 {\it Physica} C {\bf 300} 225
\bibitem{Reyes}Reyes A P, Tang X P, Bachman H N, Halperin W P, Martindale J A and Hammel P C 1997 {\it Phys. Rev.} B {\bf 55} R14737 
\bibitem{ItohTL2201}Itoh Y, Michioka C, Yoshimura K, Hayashi A and Ueda Y 2005 {\it J. Phys. Soc. Jpn.} {\bf 74} 2404 
\bibitem{2band}Suhl H, Matthias B T and Walker L R 1959 {\it Phys. Rev. Lett.} {\bf 3} 552
\bibitem{Rav}Maignan A, Putilin S N, Hardy V, Simon Ch and Raveau B 1996 {\it Physica} C {\bf 266} 173

\bibitem{ItohHg1201}Itoh Y, Machi T, Adachi S, Fukuoka A, Tanabe K and Yasuoka H 1998 {\it J. Phys. Soc. Jpn.} {\bf 67} 312
\bibitem{ItohHg1212}Itoh Y, Tokiwa-Yamamoto A, Machi T and Tanabe K 1998 {\it J. Phys. Soc. Jpn.} {\bf 67} 2212
\bibitem{Tokunaga}Tokunaga Y, Ishida K, Magishi K, Oshugi S, Zheng G q, Kitaoka Y, Asayama K, Iyo A, Tokiwa K and Ihara H 1999 {\it Physica} B {\bf 259-261} 571

\bibitem{ItohNum}Itoh Y 2015 {\it Physics Procedia} {\bf 65} 41
\bibitem{Mukuda}Mukuda H, Shimizu S, Iyo A and Kitaoka Y 2012 {\it J. Phys. Soc. Jpn.} {\bf 81} 011008
\bibitem{Mukuda5}Mukuda H, Yamaguchi Y, Shimizu S, Kitaoka Y, Shirage P and Iyo A 2008 {\it J. Phys. Soc. Jpn.} {\bf 77} 124706

\bibitem{ItohHg12010}Itoh Y, Machi T, Fukuoka A, Tanabe K and Yasuoka H 1996 {\it J. Phys. Soc. Jpn.} {\bf 65} 3751

\bibitem{Jperp}Rossat-Mignot J, Regnault L P, Jurgens  M J, Vettier C, Burlet P, Henry J Y and Lapertot G 1990 {\it Physica} B {\bf 163} 4 
\bibitem{Jperp1}Reznik D, Bourges P, Fong H F, Regnault  L P, Bossy J, Vettier C, Milius  D L, Aksay I A and Keimer B 1996 {\it Phys. Rev.} B {\bf 53} R14741
\bibitem{Jperp2}Hayden S M, Aeppli G, Perring T G, Mook H A and Do\u{g}an F 1996 {\it Phys. Rev.} B {\bf 54} R6905 

\bibitem{MM}Millis A J, Ioffe L B, Monien H 1995 {\it J. Phys. Chem. Solids} {\bf 56} 1641
\bibitem{Norman}Norman B, Kohno H and Fukuyama H 1995 {\it J. Phys. Soc. Jpn.} {\bf 64} 3903
\bibitem{JK}Kishine L 1996 {\it Preprint} arXiv:cond-mat/9604128
\bibitem{Kishine}Kishine J 1997 {\it Physica} C {\bf 282-287} 1771

\bibitem{Jurugen1}Rybicki D, Kohlrautz J, Haase J, Greven M, Zhao X, Chan M K, Dorow C J and Veit M J 2015 {\it Phys. Rev.} B {\bf 92} 081115 
\bibitem{Jurugen2}Haase J, Jurkutat M, Kohlrautz J 2017 {\it Condense. Matter} {\bf 2} 16
\bibitem{ItohHF}Itoh Y, Machi T and Yamamoto A 2018 {\it J. Phys. Conf. Series} {\bf 1054} 012006
\bibitem{Bleaney}Bleaney B, Bowers  K D and Pryce M H L 1955 {\it Proc. Roy. Soc. London}, Ser. A {\bf 228} 166

\end{thebibliography}
\end{document}